# Modified Mach-Zehnder interferometer for spatial coherence measurement

F.J. Torcal-Milla, *J. Lobera, E.M. Roche, A.M. Lopez, V. Palero, N. Andres, and M.P. Arroyo

*Grupo de Tecnologías Ópticas Láser, Instituto de Investigación en Ingeniería de Aragón (i3A), Universidad de Zaragoza, Zaragoza (Spain).*
*Corresponding author: fjtorcal@unizar.es*

**Spatial coherence of light sources is usually obtained by using the classical Young's interferometer. Despite the original experiment has been improved in successive works, some drawbacks still remain. In addition, several pairs of points must be used to characterize the totally the 2D complex-coherence degree of the source. In this work, an alternative based on a modified Mach-Zehnder interferometer which includes a pair of lenses is presented. With this modified Mach-Zehnder interferometer, we are able to measure the spatial coherence length between any two points of the laser beam section simultaneously. The set-up does not have any movable part, which makes it robust and portable. To test it, the two-dimensional spatial coherence of a high-speed laser with two cavities has been measured for different pulse energy values. We observe from the experimental measurements that the complex degree of coherence changes with the selected output energy. Both laser cavities seem to have similar complex coherence degree for maximum energy although it is not symmetrical. Thus, this analysis will allow us to determine the best configuration of the double cavity laser for interferometric applications. Furthermore, the proposed approach can be applied to any other light sources.**

Coherence represents one of the most important properties of light sources [1-3]. Some amount of coherence is necessary to produce interferences that can be used for many different applications [4, 5]. Therefore, coherence characterization of light sources is an important issue that has been investigated along the years [5-12]. Temporal and spatial coherence properties of the laser sources are usually studied separately. The first one gives the amount of correlation between fields emitted by one point of the source at different moments [9, 10]. The second one gives the correlation between the fields emitted by two different points of the source at the same time [13, 14]. The most common set-up used to measure the spatial coherence of light sources is the classical Young's interferometer, consisting of two tiny pin-holes placed after the source, whose light interferes. The visibility of the fringes is directly related to the spatial coherence between those points [4]. In practice, this set-up presents some drawbacks such as the no uniformity of the beam and the lack of enough intensity at the interference plane. Besides, it must be repeated with many pairs of points to obtain the complete beam spatial coherence degree.

In recent years, some approaches have been proposed to overcome some of the mentioned issues. For example, diffraction gratings are used in [15], fiber optics is used in [16], or a modified Mach-Zehnder interferometer is used in [17]. No one of them is able to measure the two-dimensional spatial coherence of the source, rapidly and conveniently. This handicap has been recently overcome in [18], where the authors use a mirror-based scanning wavefront-folding interferometer. In this case, the interferometer is similar to a Mach-Zehnder interferometer where two beam-splitters are substituted by two mirrors in one of the arms and the other one is enlarged by using other two extra mirrors. The extra mirrors are placed out of the plane of the interferometer and are attached to a motorized stage or piezoelectric. Thus, the authors are able to measure the two-dimensional spatial coherence, avoiding the shadows produced by the beam-splitters corners in the original configuration. The main disadvantage of the mentioned set-up would be the robustness and compactness, since the mirrors must be placed out of the interferometer plane and be attached to motorized stages. It could result mechanically less stable if we pretend to move it to another place, making the interferometer non-portable.

In this work, we introduce another alternative of modified Mach-Zehnder interferometer in which all optical elements are placed at the same plane. The introduced modifications consist of placing two lenses in one of the arms of the classical Mach-Zehnder interferometer as a confocal system. This way, the beam of one arm is inverted in both transversal directions, and the interference of both beams gives a measurement of the spatial coherence between any two points of the beam simultaneously. The modified Mach-Zehnder interferometer is shown in Fig. 1. As can be observed, the collimated beam enters the interferometer and it is directed through both arms by a 50/50 beam-splitter. Then, one of the beams is inverted in both perpendicular axes by using the confocal system, and a phase plate is placed into the other arm to compensate the phase delay introduced by both lenses. It is not necessary to match perfectly both optical paths since we are not measuring the temporal coherence of the source. It is enough to obtain interferences with measurable visibility at the observation plane. The interference fringes, Fig. 2a, are imaged by using a camera at one of the outputs. The laser used to perform the experiments is a Pegasus-PIV by NewWave (λ=527nm). It is a High-speed laser commercialized for Particle Image Velocimetry (PIV) whose coherence properties are not provided by the manufacturer. It comes with two cavities, that can be synchronized with any convenient delay and they have independent pulse energy controls. The camera is a HT-12000-S-M by Emergent Vision Technologies, whose pixel size and resolution are 3.45x3.45 µm² and 4096 x 3000 pixels, respectively.

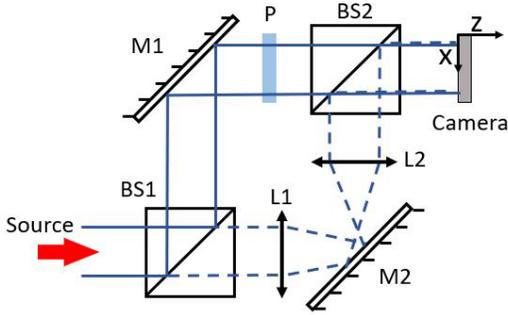

**Fig. 1.** Scheme of the modified Mach-Zehnder interferometer used to measure the spatial coherence degree of the source. M are mirrors, L are lenses, P is a phase plate, and BS are beam-splitters.

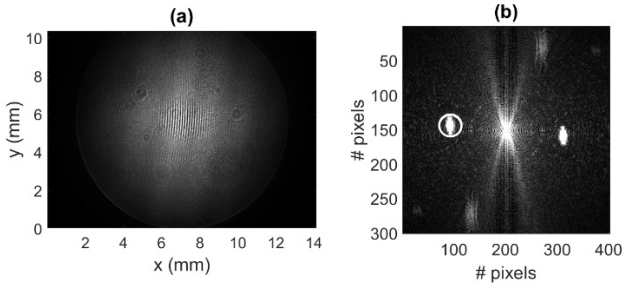

**Fig. 2.** (a) Interferogram obtained with the modified Mach-Zehnder interferometer and the laser (cavity one) at maximum pulse energy and (b) central area of the Fourier Transform of Fig.2a. The white circle marks the area cropped to isolate the first order and transform it back.

As can be observed in Fig. 2a, the interference area is centered onto the camera sensor. Some diffractive artifacts, due to dust, are also apparent, although they do not influence the measurement. The intensity at the interferogram can be expressed as

$$I = |a_1|^2 + |a_2|^2 + 2|\gamma a_1 a_2| \cos\Delta\phi, \quad (1)$$

where $a_1$ and $a_2$ are the complex amplitude of the optical fields coming from both arms of the interferometer, and $\gamma$ is the complex coherence degree. The phase difference $\Delta\phi$ is mainly due to the angle between both laser beams at each point of the sensor $\vec{r}$.

$$\Delta\phi(\vec{r}) = (\vec{k}_1 - \vec{k}_2)\vec{r} + \delta(\vec{r}), \quad (2)$$

where $\vec{k}_1$ and $\vec{k}_2$ are the vector wave of the corresponding collimated beams coming from both arms of the interferometer to any point of the camera, $\vec{r}$.

As the wave vector is constant in our specific configuration, this phase difference introduces a uniform periodicity in the horizontal direction, and the phase difference is roughly $\Delta\phi(\vec{r}) \approx 2\pi f_{ox} x$. Any discrepancy with our measured phase difference, $\delta(x,y) = \Delta\phi(x,y) - 2\pi f_{ox} x$, should be due to the fact that the light coming from each arm is not perfectly collimated and therefore, due to the spatial coherence of the beams.

In the Fourier domain we can separate the different terms of the recorded interferogram.

$$\Im(I) = \Im(|a_1|^2) + \Im(|a_2|^2) + \Im(\gamma a_1 a_2^*) + \Im(\gamma^* a_1^* a_2) \quad (3)$$

Fig. 2b shows the Fourier Transform of one typical interferogram, such as that shown in Fig. 2a. The white circle marks one of the interference terms centered around $f_{ox}$. We can isolate it, shift it to the center of the spectrum, and Fourier transform back [19]. The resultant complex value is related with the coherence degree as

$$b = \gamma a_1 a_2^* \exp(-i2\pi f_{ox} x). \quad (4)$$

Let us note that the phase difference due to the geometry of our system has been removed, so the remaining phase difference comes from the spatial decorrelation of both beams, $\delta(\vec{r})$. The modulus of $b$ can be directly related to the coherence degree. Then, if we capture the intensity of each arm, $|a_1|^2$ and $|a_2|^2$ separately, we can compute the complex degree of coherence as

$$\gamma = \frac{b}{\sqrt{|a_1|^2 |a_2|^2}} \quad (5)$$

In Fig. 3a we show the absolute value of the complex degree of coherence computed from Eq. (5) at the camera sensor in any direction within the two-dimensional section of the laser beam. For our particular light source, we can observe that coherence is longer in the horizontal direction but presents a kind of replica or lobes along the vertical direction that could be due to excitation of higher modes in the laser cavity.

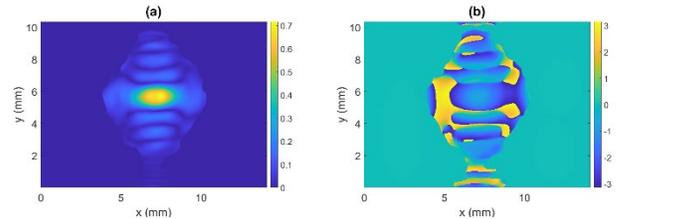

**Fig. 3.** (a) Absolute value of the two-dimensional complex degree of coherence of the laser (cavity one, maximum energy), (b) $\delta(\vec{r})$, i.e. phase of the two-dimensional complex degree of coherence of the laser (cavity one, maximum energy).

In addition, we show in Fig. 3b the phase of the complex degree of coherence that it is far from being uniform. Thus, the complex coherence degree cannot be approximated as a real valued distribution as we might need to consider its phase distribution in some interferometric applications. From Fig. 3a, it is easy to obtain the spatial coherence along the horizontal and vertical directions through the center just taking its central row and column, respectively. Fig. 4 shows these profiles. As can be seen, the coherence degree along the y-axis has areas far from the center where it increases again, allowing interferences with measurable visibility. From these profiles, we can evaluate the coherence length as the distance at which the coherence degree decreases at *1/e* of the maximum (see Fig. 4).

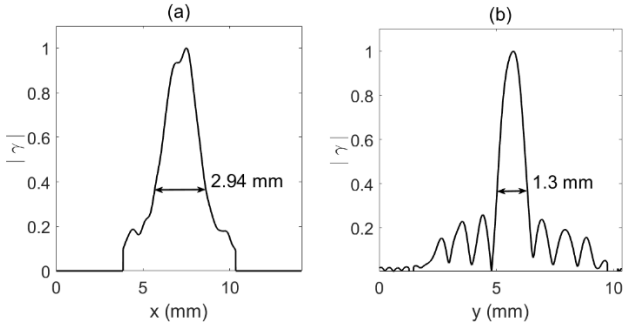

**Fig. 4.** Absolute value of the complex degree of coherence of the laser (cavity one, maximum energy) at perpendicular directions through the center, (a) along x-axis and (b) along y-axis.

Furthermore, to check the dependence of the complex degree of coherence on the laser pulse energy, we have repeated the experiment for different values of the pulse energy of the cavity one. We show in Fig. 5 the absolute value and phase of the complex degree of coherence for increasing pulse energy. The two-dimensional maps of the complex coherence degree show multiple lobes, where the beam is spatially coherent. It starts being quite symmetric for low pulse energy, with the same behavior along both perpendicular axes, and becomes more and more non-symmetric as we increase the pulse energy. On the contrary, the corresponding phase maps seem to have no significant changes with increasing energy of the laser pulse. Fig. 6 shows the horizontal and vertical central absolute value of the complex degree of coherence corresponding to Fig. 5. It has a small dependence on the pulse energy, although along the y-direction, it remains almost constant. Therefore, the coherence length along both axes almost does not depend on the pulse energy of the laser. On the other hand, some lesser powerful maxima of the coherence degree are observed at both sides of the central maximum along the y-direction. These secondary maxima appear at different positions and with different power depending on the energy per pulse of the laser.

In order to be able to compare both cavities of the laser, we have also tested the cavity two, (Fig. 7 and Fig. 8). Although both cavities should be equal, the complex degree of coherence is partially different. With the second cavity, the spatial coherence is also circularly symmetrical for low energy per pulse and becomes elongated along the x-direction for higher energy. In addition, we also observe in Fig. 8 that secondary maxima with lower coherence degree appear along the y-axis at different positions than for the first cavity of the laser.

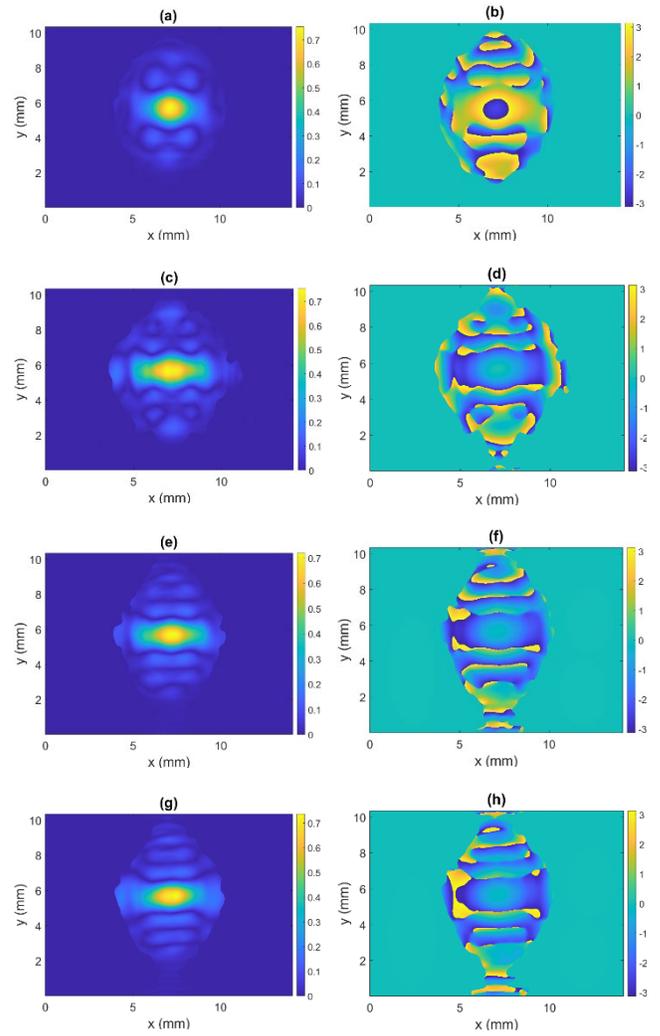

**Fig. 5.** Absolute value (left column) and $\delta(\vec{r})$, i.e. phase (right column) of the complex degree of coherence of the laser (cavity one) for different pulse energy. Top to bottom means increasing the lasing energy from 40% to 85% of the maximum value.

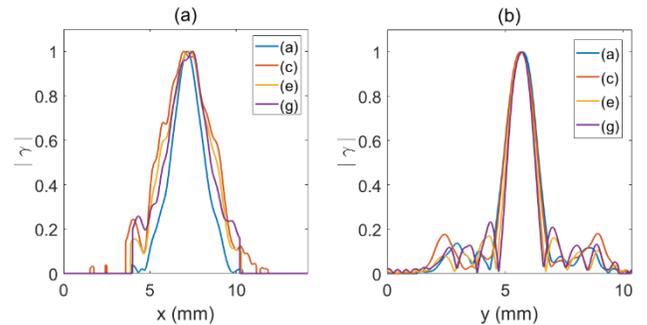

**Fig. 6.** Absolute value of the complex degree of coherence of the laser (cavity one) at perpendicular directions through the center, (a) x-axis and (b) y-axis, corresponding to the pulse energies used in Fig. 5.

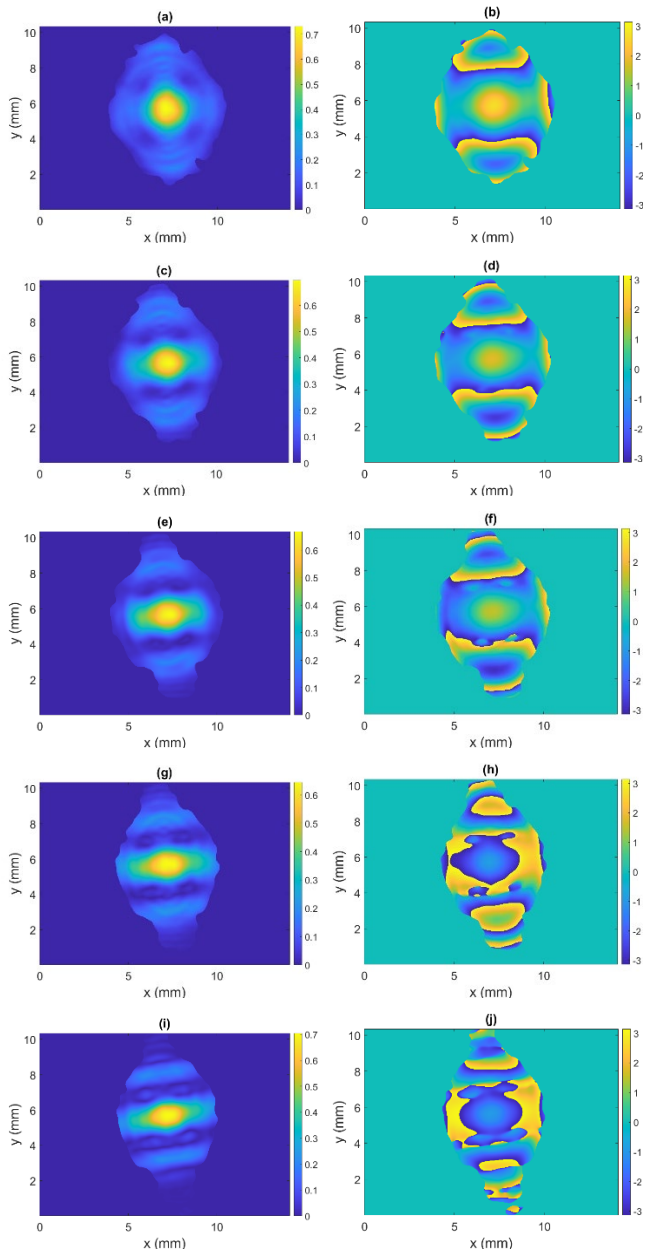

**Fig. 7.** Absolute value (left column) and $\delta(\vec{r})$, i.e. phase (right column) of the complex degree of coherence of the laser (cavity two) for different pulse energy. Top to bottom means increasing the lasing energy from 40% to 100% of the maximum value.

To conclude, in this work a modified Mach-Zehnder interferometer for spatial coherence characterization of light sources is presented. Unlike other set-ups, this one results more robust and portable, since all optical components are placed at the same plane and no motorized stages are needed. We have tested it with a high-speed laser with two cavities, commonly used for PIV applications but it can be used for any other light sources. The complex coherence degree is not symmetric for none of the cavities except when they are working at low energy per pulse. From the experimental measurements, we conclude that both cavities of the laser behave similarly. The differences appear in the secondary maxima, but we think it is not relevant for interferometric purposes.

This analysis will allow us to determine the best configuration of the double cavity laser to be used for interferometric applications. Although the degree of coherence is not symmetrical, lasers at maximum energy per pulse seem to be the better option. The coherent area is quite equal for both of them. The only thing to keep in mind is that could be better to orient the beams so that they interfere along the x-axis, since the spatial coherence is longer.

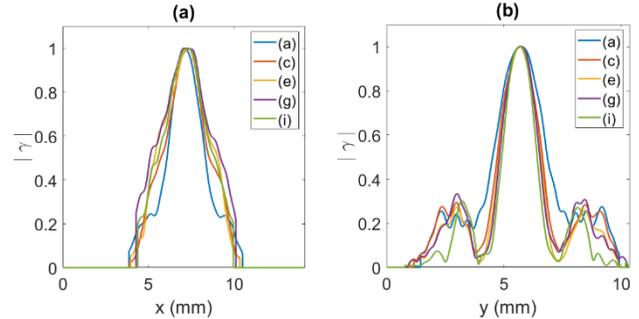

**Fig. 8.** Absolute value of the complex degree of coherence of the laser (cavity two) at perpendicular directions through the center, (a) x-axis and (b) y-axis, corresponding to the pulse energies used in Fig. 7.

**Funding.** Gobierno de Aragón-Fondo Social Europeo (Grupo de Tecnologías Ópticas Láser E44_20), and Ministerio de Ciencia e Innovación of Spain (Project PID2020-113303GB-C22).

**Disclosures.** The authors declare no conflict of interest.